\documentclass[runningheads]{svmult}

\usepackage{makeidx}   
\usepackage{graphicx}  
\usepackage{subeqnar}  
\usepackage{multicol}  
\usepackage{physprbb}  
\makeindex             

\def\gtrsim{\mathrel{\hbox{\rlap{\hbox{\lower4pt\hbox{$\sim$}}}\hbox{$>$}}}}
\begin{document}
\title*{BATSE Observations of Fast X-ray Transients Detected by
BeppoSAX-WFC}
\toctitle{BATSE Observations of Fast X-ray Transients Detected by
BeppoSAX-WFC}
%
%
\titlerunning{BATSE Observations of Fast X-ray Transients}
%
\author{R.~Marc~Kippen\inst{1,2}
\and Peter~M.~Woods\inst{1,3}
\and John~Heise\inst{4}
\and Jean~in't~Zand\inst{4}
\and Robert~D.~Preece\inst{1,2}
\and Michael~S.~Briggs\inst{1,2}}
\authorrunning{R.~M.~Kippen et al.}
\institute{National Space Sci.\ and Tech.\ Ctr., 320 Sparkman Dr.,
Huntsville, AL 35805, USA
\and University of Alabama in Huntsville, Huntsville, AL 35899, USA
\and Universities Space Research Association, Huntsville, AL 35805, USA
\and SRON-Utrecht, Sorbonnelaan 2, NL-3584 CA Utrecht, The Netherlands}

\maketitle              

\begin{abstract}
The {\it Beppo\/}SAX Wide Field Cameras have been successful in
detecting gamma-ray bursts in the 2--26 keV energy range.  While most
detected bursts are also strong emitters at higher energies, a
significant fraction have anomalously low gamma-ray flux.  The nature
of these ``Fast X-ray Transients'' (FXTs), and their relation to
gamma-ray bursts (GRBs), is unknown.  We use BATSE untriggered
continuous data to examine the $>$20 keV gamma-ray properties of the
events detected in common with {\it Beppo\/}SAX.  Temporal and
spectral characteristics, such as peak flux, fluence, duration, and
spectrum are compared to the full population of triggered BATSE GRBs.
We find that FXTs have softer spectra than most triggered bursts, but
that they are consistent with the extrapolated hardness expected for
low-intensity GRBs.
\end{abstract}

\section{Introduction}
Several papers in these proceedings detail how measurements made with
the {\it Beppo\/}SAX Wide Field Cameras (WFCs) have revolutionized
gamma-ray burst (GRB) astronomy in two key areas: First, the timely,
arc-minute burst locations provided by the WFCs led directly to
ground-breaking discoveries of multi-wavelength afterglow emission and
host galaxies at cosmological distances.  Second, the 2--26 keV WFC
burst measurements extend GRB spectroscopy into the x-ray band, where
few previous observations exist.  This latter capability has led to
the exciting discovery of several ``Fast X-ray Transients'' (FXTs),
which resemble GRBs in their x-ray properties and spatial
distribution, but lack the strong gamma-ray emission typical of
``classical'' GRBs \cite{heise}.  Some bursts with a strong x-ray
spectral component were observed previously with {\it Ginga\/}
\cite{stroh}, hinting that the WFC FXTs could represent a class of
x-ray rich GRBs.  However, a completely unrelated and unknown
phenomenon cannot be ruled out.

The key to understanding the relation between FXTs and GRBs lies in
the gamma-ray domain, where we have the most knowledge of classical
GRB behavior to compare with.  Unfortunately, observational data are
scarce due to the weakness of the gamma-ray emission from these
events.  In this paper, we present preliminary results on the
gamma-ray properties of WFC FXTs compared to those of classical GRBs
using data from the {\it Compton\/}-BATSE Large Area Detectors (LADs,
$>$20 keV).  BATSE is the only instrument to detect significant
gamma-ray emission from WFC FXTs, and thus provides the best means of
understanding their true nature.

\section{BATSE Observations}
BATSE and {\it Beppo\/}SAX-WFC operated simultaneously for 3.8 years,
ending with the termination of the {\it Compton\/} Observatory on 4
June, 2000.  In this interval, $\sim$53 GRB-like transient events were
observed by the WFCs -- 17 of which were classified as FXTs due to
their lack of detectable gamma-ray emission in the SAX GRB Monitor
(40--400~keV).  Based on the WFC source locations, we know that 12 of
these FXTs were observable (unocculted by Earth) by BATSE, but none
activated BATSE's on-board transient event trigger system.  This
indicates that the 50--300~keV peak fluxes were near, or below
$\sim$0.2~$\rm{ph} \cdot \rm{cm}^{-2} \cdot \rm{s}^{-1}$ (1024 ms
timescale).

\begin{figure}[h]
\begin{center}
\includegraphics[height=2.00in]{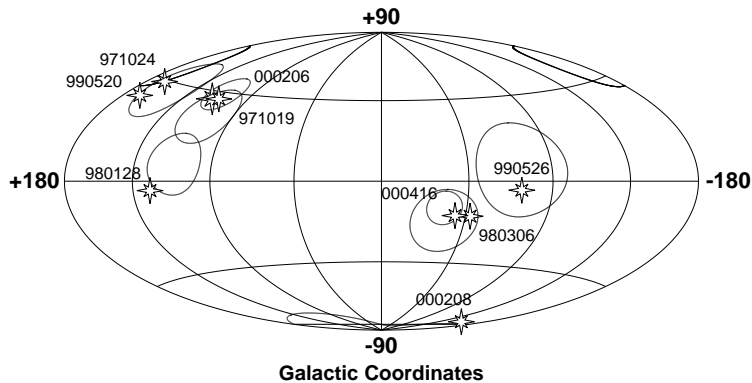}
\end{center}
\vspace{-0.4cm}
\caption[]{Locations (and dates) of the nine WFC FXTs detected with BATSE 
in this study.  The BATSE 1$\sigma$ error circles (solid lines) are compared
to the independent WFC locations (stars whose size is not representative
of the $\sim$10 arc-minute uncertainties).}
\label{fig1}
\end{figure}

To investigate with higher sensitivity than the on-board trigger
allows, we performed a {\it post facto\/} search for the 10 FXTs where
BATSE continuous data are available.  Nine of the 10 candidates were
detected with $\geq$5$\sigma$ significance in the 20--100~keV energy
range.  These detections are strengthened by the fact that the times
and independent BATSE locations of the events agree (within the
uncertainties) with those measured with the WFCs (see Figure
\ref{fig1}).

\vspace{-0.3in}

\section{Gamma-Ray Properties of FXTs Compared to GRBs}
Qualitatively, the nine FXTs detected in the BATSE search are very
similar to GRBs in their temporal and spectral characteristics.  For
instance, they have rapidly varying lightcurves, durations from
$\sim$1--50 s, and strong spectral evolution.  To quantify the
comparison, 4-channel, 1024-ms data were used to compute several
standard GRB parameters, including peak flux, fluence, duration, and
hardness ratios.  These quantities were produced using the same
software and procedures employed for all BATSE GRBs, so we can make
direct comparisons to the large BATSE GRB trigger catalog \cite{bcat}.

\begin{figure}[t]
\begin{center}
\includegraphics[width=1.0\textwidth]{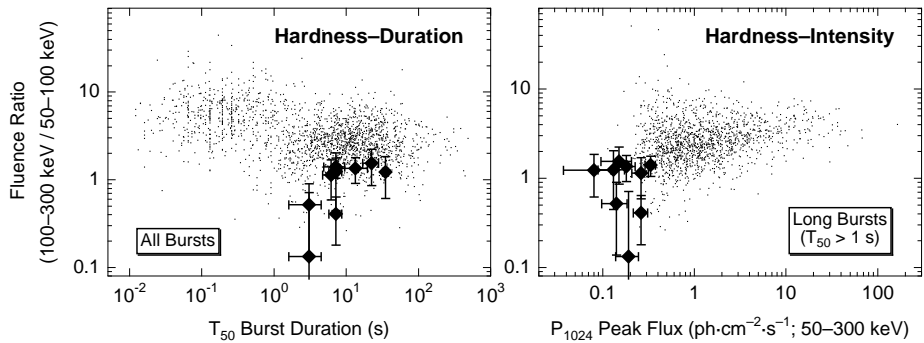}
\end{center}
\vspace{-0.4cm}
\caption[]{Gamma-ray parameters of nine FXTs (diamonds) compared to 
GRBs from the BATSE trigger catalog (dots).  The plot at right shows 
only ``long'' GRBs (T$_{50} > 1$ s).}
\label{fig2}
\end{figure}

As can be seen in Figure \ref{fig2} (left), the FXTs have durations
comparable to the long class (T$_{50} > 1$ s) of GRBs, but are
noticeably softer than the average long GRB based on their fluence
hardness ratios.  This spectral difference is also apparent in the
distribution of peak flux hardness ratios.  However, since the FXTs
are less intense than most triggered GRBs, spectral differences may be
expected due to the well-known GRB hardness--intensity correlation
(e.g., \cite{mallo}).  As shown in Figure \ref{fig2} (right), the FXTs
appear to be generally consistent with the extrapolated
hardness--intensity trend of long-duration bursts (albeit three of the
FXTs are anomalously soft).  This result offers tantalizing evidence
that the FXTs could be a natural extension of known GRB
characteristics.

\begin{figure}[t]
   \begin{minipage}[t]{0.48\textwidth}
      \begin{center}
         \includegraphics[height=2.2in]{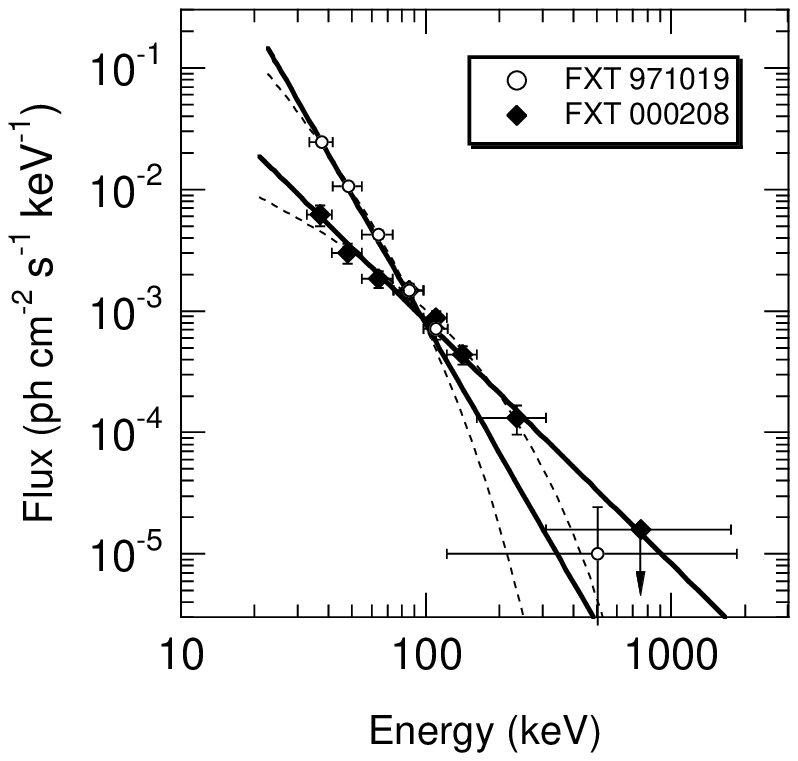}
      \end{center}
      \vspace{-0.4cm}
      \caption{Time-averaged energy spectra of two FXTs measured with
      BATSE.  Solid and dashed lines indicate the best-fit power-law
      and Comptonized models, respectively.}
      \label{fig3}
   \end{minipage}
   \hfill
   \begin{minipage}[t]{0.48\textwidth}
      \begin{center}
         \includegraphics[height=2.2in]{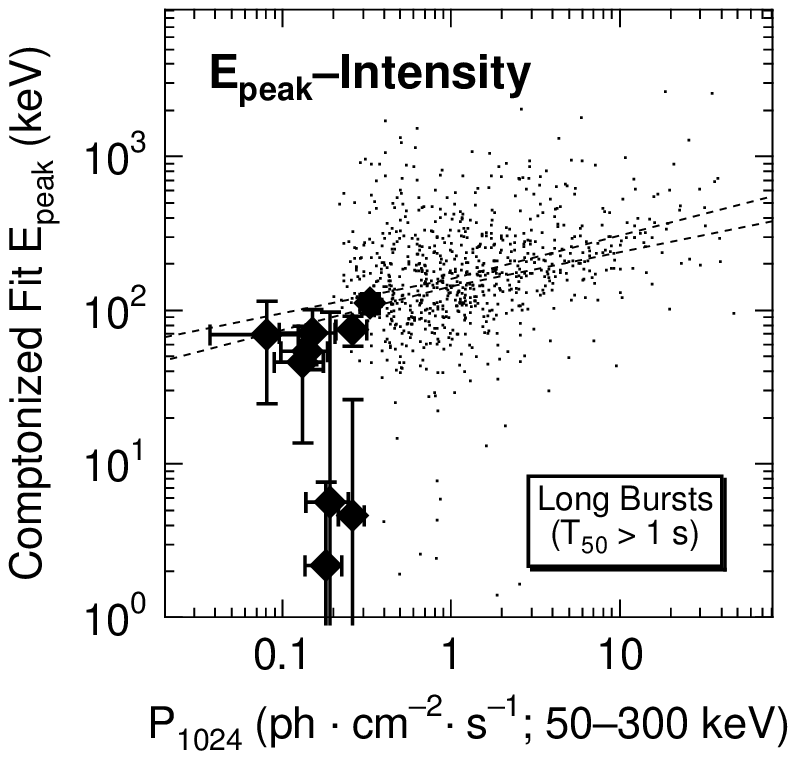}
      \end{center}
      \vspace{-0.4cm}
      \caption{Best-fit time-averaged $E_{\rm{peak}}$ vs.\ peak flux for FXTs
      (diamonds) and long GRBs (dots).  Dashed lines indicate the 68\%
      confidence region of a power-law fit to the GRB
      $E_{\rm{peak}}$--intensity correlation.}
      \label{fig4}
   \end{minipage}
\end{figure}

To investigate this result in more detail, 16-channel, 2.048-s data
were used to fit the time-averaged gamma-ray spectrum of each FXT.
Due to the large statistical uncertainties in these weak events, only
simple spectral models with few free parameters are justified.  We
find that both the single power law model and the ``Comptonized''
model \cite{mallo} yield acceptable fits to the data.  Figure
\ref{fig3} shows examples of two FXT spectra and their best-fit
spectral models.  

The Comptonized model, which includes curvature parameterized by the
$\nu \cal{F}_{\nu}$ peak energy $E_{\rm{peak}}$, was used by Mallozzi
et al.\ \cite{mall2} to fit all time averaged 16-channel spectra of
1023 bursts from the BATSE 4B catalog.  In Figure \ref{fig4}, the
fitted values of $E_{\rm{peak}}$ for 802 long GRBs from this
collection are compared to those of the FXTs. Apart from the three
soft outliers mentioned above (note also the large errors), the FXTs
appear to be consistent with $E_{\rm{peak}}$--intensity trend of the
GRB sample.  For further clarity, this trend was modeled by fitting a
power-law function to the distribution of $E_{\rm{peak}}$ in eight
intensity bins (each containing $\sim$100 long GRBs).  The 68\%
confidence region for this fit is indicated by the two dashed lines in
Figure \ref{fig4}.  The extrapolated fit is clearly consistent with
most of the FXT spectra, in agreement with the previous result based
on hardness ratios.

\section{Conclusions and Future Work}
Thanks to the high sensitivity of BATSE, we have been able to directly
compare the gamma-ray properties of FXTs to those of the full GRB
population.  The preliminary result is that FXTs appear to be
consistent with an extrapolation of known GRB behavior -- indicating
that they represent a previously unexplored sector of the GRB
population.  If confirmed, this result implies that there are a large
number of undetected bursts with peak energies in the x-ray regime.
In fact, the untriggered BATSE burst catalog of Stern et al.\
\cite{stern} suggests that the all-sky rate of FXT-like events with
$P_{1024} \gtrsim 0.1$~$\rm{ph} \cdot \rm{cm}^{-2} \cdot \rm{s}^{-1}$
is $\sim$400/yr.  This represents nearly half of all long GRBs in
their catalog!  It is tempting to speculate that these x-ray rich
events could represent a large population high-redshift GRBs.  In
future work, we will attempt to confirm the findings presented here by
jointly analyzing the x-ray and gamma-ray spectral data from WFC and
BATSE.  We will also investigate in more detail the similarities
between the WFC FXTs and the large sample of untriggered BATSE events.


\begin{thebibliography}{8.}
\addcontentsline{toc}{section}{References}

\bibitem{heise} J.~Heise, J.~in't~Zand, R.~M.~Kippen, et al.: these proceedings

\bibitem{stroh} T.~E.~Strohmayer, E.~E.~Fenimore, T.~Murakami, et al.: 
ApJ {\bf 500}, 873 (1998)

\bibitem{bcat} \texttt{http://gammaray.msfc.nasa.gov/batse/grb/catalog/current/}

\bibitem{mallo} R.~S.~Mallozzi, W.~S.~Paciesas, G.~N.~Pendleton, et al.: 
ApJ {\bf 454}, 597 (1995)

\bibitem{mall2} R.~S.~Mallozzi, G.~N.~Pendleton, W.~S.~Paciesas, et al.: 
In {\it Gamma-Ray Bursts: 4th Huntsville Symp.\/} eds. C.~Meegan et al.\
(AIP CP428, New York 1998) p.\ 273

\bibitem{stern} B.~E.~Stern, Y.~Tikhomirova, et al.: ApJ, in press
(astro-ph/0009447)

\end{thebibliography}
\end{document}